\documentclass{emulateapj}

\renewcommand{\vec}[1]{\mbox{\boldmath{$#1$}}}

\shorttitle{Magnetic Field Signatures}
\shortauthors{Baldner et al.}

\begin{document}
\title{Solar Magnetic Field Signatures in Helioseismic Splitting Coefficients}
\author{Charles S.~Baldner}
\affil{Department of Astronomy, Yale University, P.O. Box 208101, New Haven, CT, 06520-8101}
\email{charles.baldner@yale.edu}
\author{H.~M.~Antia}
\affil{Tata Institute of Fundamental Research, Homi Bhabha Road, Mumbai 400005, India}
\author{Sarbani Basu}
\affil{Department of Astronomy, Yale University, P.O. Box 208101, New Haven, CT, 06520-8101}
\and
\author{Timothy P.~Larson}
\affil{Hansen Experimental Physics Laboratory, Stanford University, Stanford, CA 94305-4085}

\begin{abstract}
Normal modes of oscillation of the Sun are useful probes of the solar 
interior.  In this work, we use the even-order splitting coefficients to 
study the evolution of magnetic fields in the convection zone over solar 
cycle 23, assuming that the frequency splitting is only due to rotation
and a large scale magnetic field.  We find that the data are best fit by 
a combination of a poloidal field and a double-peaked near-surface toroidal field. 
 The toroidal fields are centered at $r_0=0.999R_\odot$ 
and $r=0.996R_\odot$ and are confined to the near-surface layers.  The poloidal 
field is a dipole field.  The peak strength of the poloidal field is $124\pm17$~G.  
The toroidal field peaks at $380\pm30$~G and $1.4\pm 0.2$~kG for the shallower 
and deeper fields respectively.  The field strengths are 
highly correlated with surface activity.  The toroidal field strength shows a 
hysteresis-like effect when compared to the global 10.7 cm radio flux.  The poloidal 
field strength shows evidence of saturation at high activity.
\end{abstract}

\keywords{Sun: activity, Sun: helioseismology, Sun: magnetic fields}

\section{Introduction}
Understanding the nature of the Sun's magnetic fields --- their structure and 
variability, their generation mechanisms, and their effects on the heliosphere 
--- is one of the key aims of current research in solar physics.  It is generally believed 
that the magnetic fields are generated by a cyclic dynamo that operates somewhere 
in the solar interior.  In this paper, we use helioseismology to study the 
global scale internal magnetic fields over the course of solar cycle 23.  

Helioseismology is the most powerful tool available to solar physicists to 
study the interior of the Sun.  The oscillation frequencies have been used to 
study the structure and dynamics of the solar interior with great precision.  
Magnetic fields, however, have proved to be much more challenging.  There are a
number of important difficulties in dealing with magnetic fields in a 
helioseismic context.  The magnitudes of the signatures in the data are quite 
small, making statistically significant measurements challenging.  Secondly, 
the interpretation of data is very difficult.  The physics of wave propagation 
in the presence of magnetic fields is far more complex than in the non-magnetic
case.  Further, the geometry of the underlying field strongly affects 
the signatures in helioseismic global mode frequencies, meaning different 
field configurations and strengths can be difficult to distinguish from their 
helioseismic signatures.  Even worse, \citet{ZG95} showed that because 
magnetic fields act on mode frequencies both by perturbing the thermal 
structure of the Sun and by changing the wave propagation speeds directly, 
there is a degeneracy between magnetic field effects and other thermal 
perturbations which cannot be distinguished a priori from helioseismic data.

Although helioseismic determinations of magnetic fields are difficult, 
there have been many attempts to do so.  \citet{Isaak82} suggested that 
the then observed frequency splittings in the solar acoustic spectrum 
could be caused by a large scale magnetic field situated in the core.  
\citet{DG84} used an asymptotic approximation to study the effects of 
magnetic field on the splitting coefficients, and \citet{DG88} 
argued that a 1 MegaGauss field at the base of the convection zone was 
necessary to explain the observed splitting coefficients.  However, 
\citet{Basu97, ACT00} placed a limit of 0.3~MG on the field at the base of the 
convection zone; thus the situation was unclear.  A mega-Gauss magnetic field
is also inconsistent with dynamo theories and constraints from
other observations \citep[e.g.,][]{DC93}.

\citet{GT90} developed a formalism to compute the effects of rotation and 
axisymmetric magnetic fields on the frequency splittings (discussed in the 
following section), which \citet{ACT00} used to analyze the first year of 
Michelson Doppler Imager (MDI) data.  They placed limits on the strengths of 
internal toroidal fields, finding a limit of 20~kG at a depth of 30~Mm, and a 
limit of 300~kG at the base of the convection zone ($r = 0.713R_\odot$).  
\citet{DGKS00} inverted the mean frequencies and splitting coefficients for 
changes in temperature, and found that the resulting temperature perturbation 
could be explained by a change in magnetic field of 60~kG at a depth of 45~Mm 
($r\sim0.93R_\odot$).

\citet{DGS01} found that changes in $f$-mode frequencies from solar minimum 
to solar maximum implied a decrease in solar radius with activity, which 
they associated with a change between 4 and 8~Mm in depth.  In explaining 
this result with changing magnetic fields, they assumed a tangled field, 
but even so the magnitude of the change in field strength was strongly 
dependent on the radial distribution of the field.  The change they 
required was 7~kG for a uniform field, or substantially less (1~kG at 8~Mm) 
for an inwardly increasing field.  \citet{CS02,CS05} looked for 
signatures of a change at the base of the convection zone from low 
activity to high activity, and found signs of a small change, which they 
proposed could be due to a change in magnetic field of 170 -- 290~kG.  
\citet{BB08}, working with an entire solar cycle's worth of helioseismic 
data, found a change in sound speed between solar maximum and solar 
minimum at the base of the convection zone, which, if due to a change in 
magnetic field, could indicate a change in field strength of 290~kG at 
that depth.

In this work, we exploit the fact that we have much  more helioseismic 
data than previous investigators had access to, and try to get a 
coherent picture of sub-surface solar magnetic fields and their
temporal evolution.  We extend 
the work of \citet{ACT00}, who considered toroidal magnetic fields, to 
include poloidal fields.  This means that we can, in principle, consider 
any axisymmetric magnetic field configuration.  We compute the effects of 
a wide variety of magnetic field configurations on the $a_2$ splitting 
coefficients, and compare them to a solar cycle's worth of MDI data.
It is not clear if the solar magnetic field has large scale structure of
the form we assume or whether it is in tangled state due to turbulence
in the convection zone. Since the effect of magnetic field manifests
through a quadratic term in magnetic field, our estimate may also be
applicable to tangled field with some degree of approximation.

\section{Perturbations to solar oscillation frequencies}
The frequencies of normal modes of oscillation $\nu_{n\ell m}$ are degenerate in $m$ in 
the case of a spherically symmetric star.  Departures from spherical symmetry 
lift this degeneracy.  When the departures from spherical symmetry are small, 
as they are in the case of the Sun, the differences in frequency for different 
values of $m$ will be small, and it is natural therefore to express the 
normal mode frequencies in terms of the mean frequency of the multiplet $\nu_{n\ell}$ 
and splitting coefficients $a_j$:
\begin{equation}
\nu_{n\ell m}
= \nu_{n\ell} + \sum_{j=1}^{j_{\rm max}} a_j (n,\ell) \, {\cal P}_j^{(\ell)}(m).
\label{acoefs}
\end{equation}
As is common in the current literature, the polynomials ${\cal P}_j^{(\ell)}(m)$ 
are the Ritzwoller-Lavely formulation of the Clebsch-Gordan expansion \citep{RL91}.  
The odd-order splitting coefficients are caused by the rotation of the Sun, and 
will not be directly considered in this work.  The even-order coefficients are 
caused by second order effects of rotation, and by the effects of magnetic 
fields or any other departure from spherical symmetry in the solar structure.
In this work, we treat rotation and magnetic fields as perturbations on the 
spherically symmetric case, which allows us to avoid explicitly constructing 
a model of a rotating, magnetized star.  The formalism was developed by \citet{GT90} 
and \citet{ACT00} extended the formalism to include the
perturbation to the gravitational potential (i.e., to relax the Cowling approximation) and to 
include differential rotation.

The first order correction to the mode frequencies due to rotation affects 
only the odd-order splitting coefficients.  These effects are due to the 
perturbation of the mode frequencies by advection of the waves.   The 
second order correction affects only the even-order splitting coefficients, 
and is caused by the perturbation to the eigenfunctions and the centrifugal 
force.  The odd-order 
coefficients can be used to determine the rotation profile $\Omega(r)$ 
\citep{Thompsonetal96,Schouetal98}, which can in turn be used to compute 
the second order rotation correction \citep{ACT00} to the even-order 
coefficients.  This correction needs to be made if the magnetic 
perturbation is comparable in size to second order rotation effect, 
which appears to be the case \citep{GT90,ACT00}.

In this work, we consider two different axisymmetric magnetic field 
configurations:  toroidal and poloidal. Following \citet{GT90}, the
toroidal field is expressed in the form
\begin{equation}\label{tor}
\vec{B} = \left[ 0, 0, a(r)\frac{d}{d\theta}P_k (\cos \theta)\right],
\end{equation}
where $P_k$ are the Legendre polynomials of degree $k$ and $a(r)$ describes the radial 
profile of the magnetic field. We consider only even values of $k$ to
ensure antisymmetry about the equator, consistent with the observed field
at the surface.  The poloidal field is assumed to be of the form
\begin{equation}\label{pol}
\vec{B} = \left[ k(k+1) \frac{b(r)}{r^2}P_k(\cos \theta), \frac{1}{r} \frac{db}{dr}\frac{d}{d\theta} P_k ( \cos \theta), 0\right],
\end{equation}
where $b(r)$ describes the radial profile of the magnetic field.
In this case we use only odd values of $k$ to ensure that the field is
antisymmetric about the equator.  With appropriate combinations of these 
two fields we can, in principal, represent any axisymmetric magnetic field.

The effect of these magnetic field configurations on the frequency splittings
of $p$-modes is calculated using the formulation of \citet{GT90, ACT00}.
There are two ways in which the magnetic field can affect the frequencies,
one is the so-called direct effect due to the additional force, and the
second is the distortion effect due to the equilibrium state being distorted 
from the original spherically symmetric one. Both these effects are included 
in all calculations.  These formulations treat the effect of these magnetic 
fields separately.  Unfortunately, the effect of magnetic fields is not linear 
and hence strictly the contributions from two different configurations cannot 
be added.  In principle, there will be some cross-terms when the combination 
of toroidal and poloidal fields have a region of overlap in the solar interior.
In this work, we neglect these terms and add the contributions from toroidal 
and poloidal fields to get the total effect. We expect the cross terms to be 
small.

\section{Data}
The data we use for comparison are 72-day mode parameter sets from the 
Michelson Doppler Imager (MDI) on the SOlar and Heliospheric Observatory 
(SOHO).  We use mode parameter sets from the corrected pipeline described 
by \citet{LarsonSchou08}.  The original MDI analysis pipeline \citep{Schou99} 
did not take in to account a number of instrumental effects which 
introduced secular trends in the mode parameter sets.  In particular, 
the plate scale of the MDI instrument has changed somewhat over SOHO's 
mission, and this results in an apparent change in the solar radius if 
not properly corrected in the analysis.  \citet{BB08} found a signature 
in the mean frequencies which became increasingly significant over the 
course of the solar cycle.  A repetition of that work with reanalyzed 
mode parameter sets removed this effect completely \citep{BBL09}.  The 
splitting coefficients, which we focus on in this work, suffer from 
similar instrumental effects as the mean frequencies, and hence we use the 
reanalyzed data in this work.

We include 56 mode sets which cover solar cycle 23.  The mode sets are 
identified by the MDI start day, beginning with set \#1216 (start day 1 May 1996), 
and ending with set \#5320 (start day 27 July 2007).  The coverage begins and 
ends at low activity, with a 10.7~cm radio flux of 72.7~SFU for the first 
set and a flux of 69.1~SFU for the last set.  The highest activity set, 
\#3160 (start day 27 August 2001), has a 10.7~cm flux of 223.9~SFU.

We fit only the $a_2$ splitting coefficients, as the higher order 
splitting coefficients have larger errors, and as such did not distinguish 
well between different field configurations.  The rotation profile 
determined from the odd-order splitting coefficients \citep{ABC08} was used to 
calculate the second-order contribution to the even-order coefficients, 
and this contribution was subtracted from the data.

\section{Results}
\subsection{Models}
\begin{figure}
\plotone{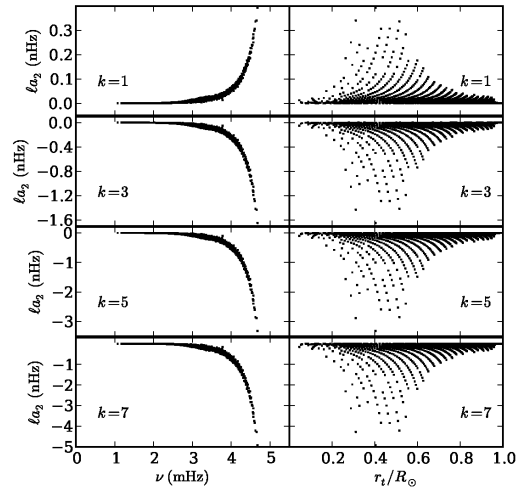}
\caption{Splitting coefficients $\ell a_2$ due to poloidal
magnetic fields.  The left hand panels are shown as a function of frequency 
$\nu$, the right hand panels are shown as a function of lower turning radius 
$r_t$.  The four configurations shown have peak field strengths of 1~G at the
surface.  The fields have four different values of $k$.  To facilitate direct 
comparison with later figures, only modes measured in the MDI data (specifically, 
the high activity set \#3160) are plotted.
\label{fig:model_pol}}
\end{figure}

In Fig.~\ref{fig:model_pol}, we show the second splitting 
coefficient for four different poloidal field configurations.  The actual 
quantities plotted are $\ell a_2$, both as a function of frequency $\nu$ 
and as a function of the lower turning radius of the modes, $r_t$.  The radial 
profile in this case is taken to be
\begin{equation}
b(r) = B_0 r^{-k},
\end{equation}
where $B_0$ is a constant which determines the peak field strength and $r$ 
is the radial distance measured in units of solar radius.  The models shown 
in Fig.~\ref{fig:model_pol} all have a peak strength of $B=1$~G (note that 
they do not all have the same value of $B_0$).  The most obvious difference 
between different order poloidal fields is that for the $k=1$ field the splitting 
coefficients are all positive, whereas for the higher order fields they are 
largely negative, although the shallow modes have positive $a_2$.

\begin{figure*}
\plotone{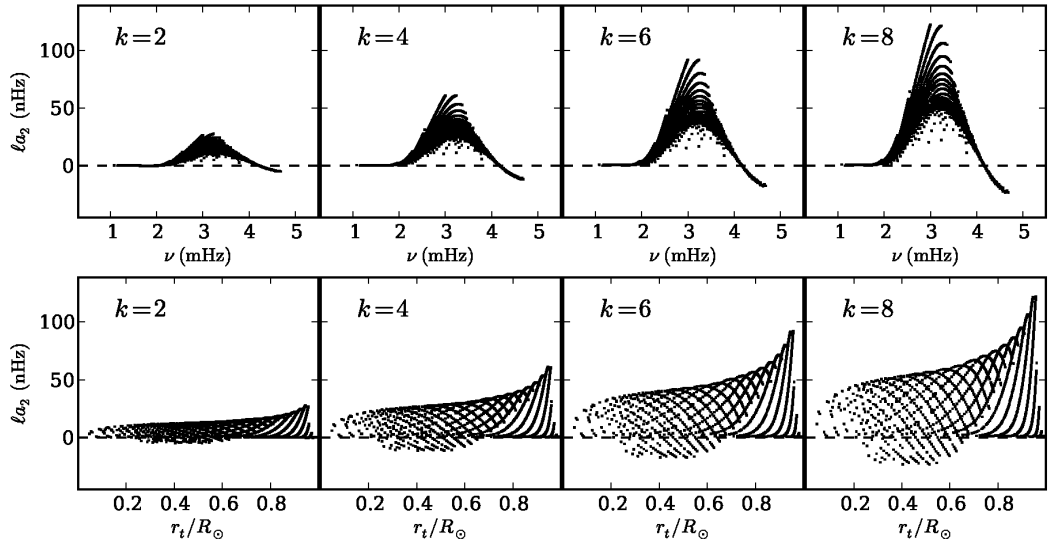}
\caption{Splitting coefficients $\ell a_2$ due to toroidal field with different 
latitudinal distributions.  The upper panels show the coefficients as a 
function of frequency $\nu$, the lower panels show the coefficients as a 
function of lower turning radius $r_t$.  All the results are with $\beta_0=10^{-4}$,
$r_0=0.999R_\odot$ and $d_0=0.001R_\odot$.  Only modes present in the MDI data 
have been plotted.
\label{fig:model_k}}
\end{figure*}

The toroidal field we employ is similar to that used by \citet{ACT00}, 
with a radial profile given by
\begin{equation}
a(r) = \left\{
\begin{array}{ll}
\sqrt{8\pi p \beta_0}\left(1-(\frac{r-r_0}{d_0})^2\right) & \textrm{if }|r-r_0|\le d_0\\
0 & \textrm{otherwise}
\end{array}
\right.
\end{equation}
where $p$ is the gas pressure, $\beta_0$ is the ratio of the 
magnetic to gas pressure at $r_0$, and $r_0$ and $d_0$ are position and
width of the field.  
As is the case for the poloidal fields, the toroidal field corrections 
are linear in magnetic field strength squared.  Excepting field strength, 
therefore, our toroidal fields are described by three quantities:  the 
order of the Legendre polynomial $k$, which determines the latitudinal 
distribution of the field, the central radius $r_0$, which determines 
the location, and the width $d_0$.  Figure \ref{fig:model_k} shows the 
splitting coefficients due to toroidal fields with different values of 
$k$ but the same radial profile (in this case, $\beta_0=10^{-4}$, 
$r_0=0.999R_\odot$ and $d_0=0.001R_\odot$).  For the $a_2$ coefficient, 
the order $k$ of the field makes very little difference except to the 
scale of the perturbation --- increasing $k$ for the same $\beta_0$ 
effectively increases the total amount of flux, but except for this 
effect, the $a_2$ coefficients are not sensitive to different latitudinal 
distributions.  For the remainder of the work, therefore, we restrict 
ourselves to $k=2$ fields.

\begin{figure*}
\plotone{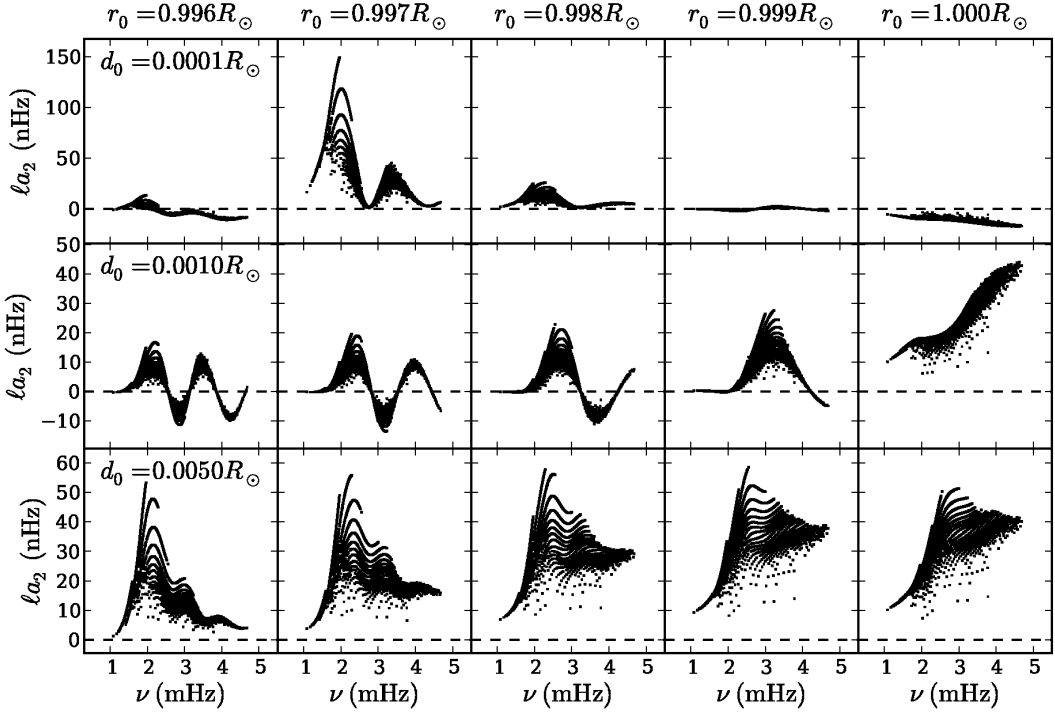}
\caption{Splitting coefficients $\ell a_2$ due to near-surface toroidal 
magnetic fields, as a function of frequency $\nu$.  The results are shown 
for $k=2$ with five different values of central radius $r_0$ (from $0.996R_\odot$ 
to $R_\odot$), and three different values of the width of the field $d_0$ 
$(0.0001, 0.001,\hbox{ and }0.005)R_\odot$.  Only modes present in the MDI 
data have been plotted.
\label{fig:model_grid_nu}}
\end{figure*}

\begin{figure*}
\plotone{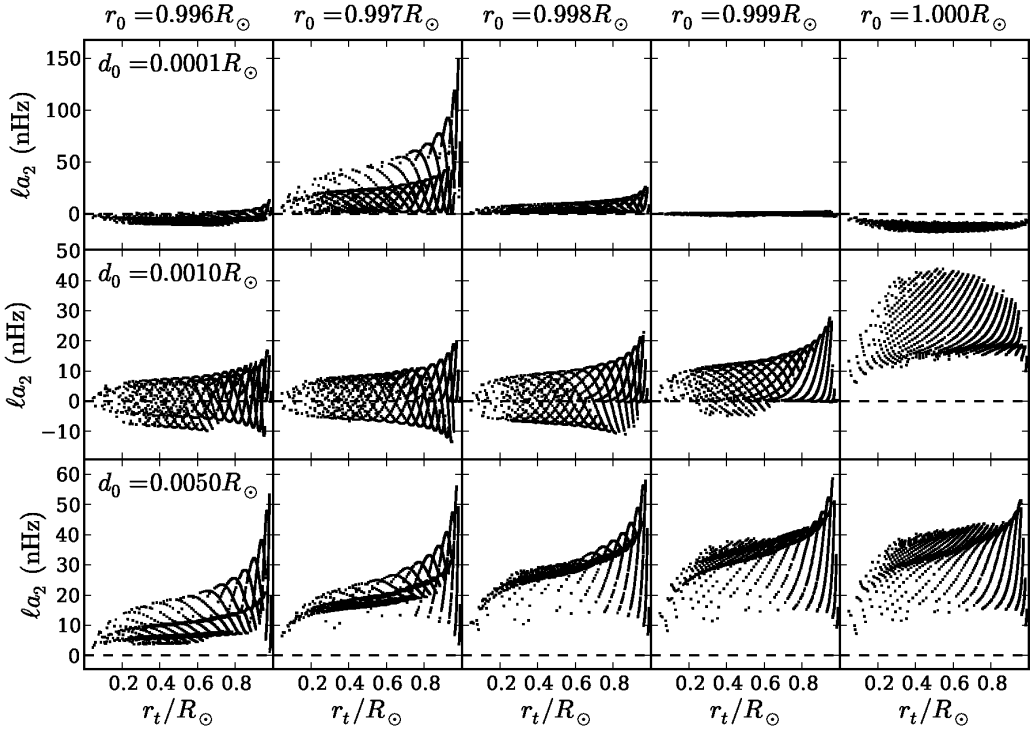}
\caption{Same as Fig.~\ref{fig:model_grid_nu}, but plotted as a function of
the lower turning radius $r_t$.
\label{fig:model_grid_rt}}
\end{figure*}

Figure \ref{fig:model_grid_nu} shows the splitting coefficients 
$\ell a_2$ for near-surface toroidal fields with different central radii $r_0$ and 
widths $d_0$ as a function of frequency.  Figure \ref{fig:model_grid_rt} 
shows the same, but as a function of the lower turning radius, $r_t$.  
The behavior of the splitting coefficients is not surprising.  
In general, the fields which penetrate below the surface 
show oscillatory behavior as a function of frequency similar to that seen in 
mode frequency \citep{G1990,GT90} and used by \citet{RV94,BAN94} and others to 
study the convection zone base.  The period of these 
oscillations is related to the acoustic depth of the perturbation in the structure. 
 Decreasing the depth of the perturbation lengthens the period 
of the oscillatory behavior.  Fields which are confined near the surface, 
on the other hand, do not exhibit oscillatory behavior, but instead resemble 
the `surface term' correction which is removed in structure inversions 
\citep[e.g.,][]{Dzetal90,AB94}.  Increasing the width of the perturbation 
smears out the oscillatory signature, as seen in Fig.~\ref{fig:model_grid_nu}.  
Because all the modes sampled have lower turning radii below the magnetic 
fields considered here, there are no obvious signatures in the splitting
coefficients as a function of $r_t$.

\begin{figure}
\plotone{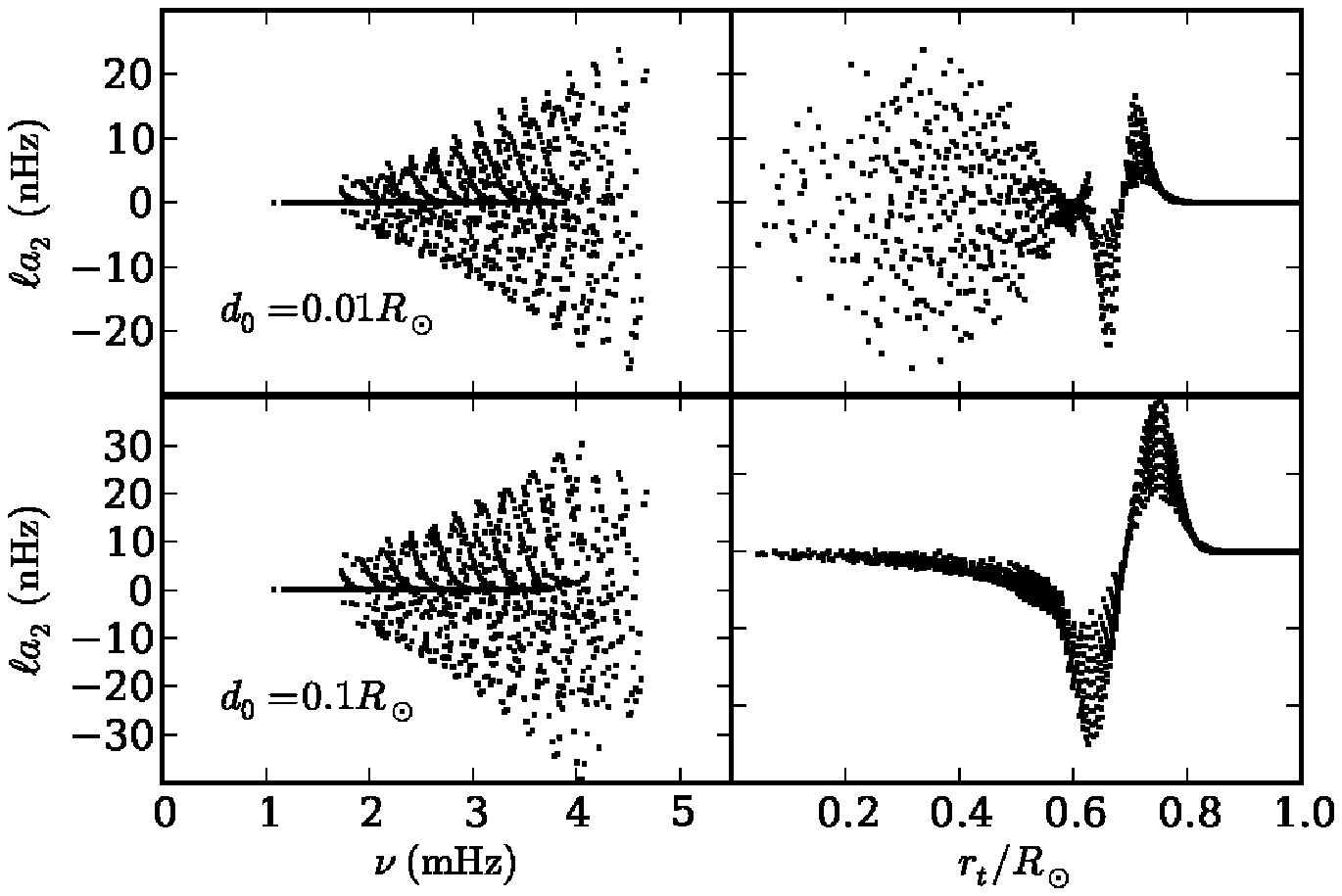}
\caption{Effects of toroidal magnetic fields at the base of the convection 
zone on the $\ell a_2$ splitting coefficients. Results for two magnetic field 
configuration with $k=2$, $r_0=0.71R_\odot$ and $\beta_0=10^{-4}$ are shown.  
The fields have widths of $d_0=0.01R_\odot$ and $d_0=0.1R_\odot$ for the top and bottom 
panels, respectively.  Left hand panels show the splitting coefficients as a 
function of frequency $\nu$, right hand panels show the splitting coefficients 
as a function of lower turning radius $r_t$.  Only observed modes have been plotted.
\label{fig:model_czb}}
\end{figure}

In addition to fields near the surface, in Fig.~\ref{fig:model_czb}, we show 
the splitting coefficients due to some toroidal fields located at the base of 
the convection zone.  The fields shown differ only in the width $d_0$ 
of the fields.  Unlike the surface fields shown in previous figures, the deep 
field signatures show both positive and negative splitting coefficients. These
models are most interesting as a function of lower turning radius $r_t$.  The 
splitting coefficients are positive above the center of the magnetic field, 
and negative below the center of the magnetic field.  Further, as the width 
of the field is increased, the width of the perturbations to the splitting 
coefficients (in $r_t$ figure) increases as well.

\begin{figure}
\plotone{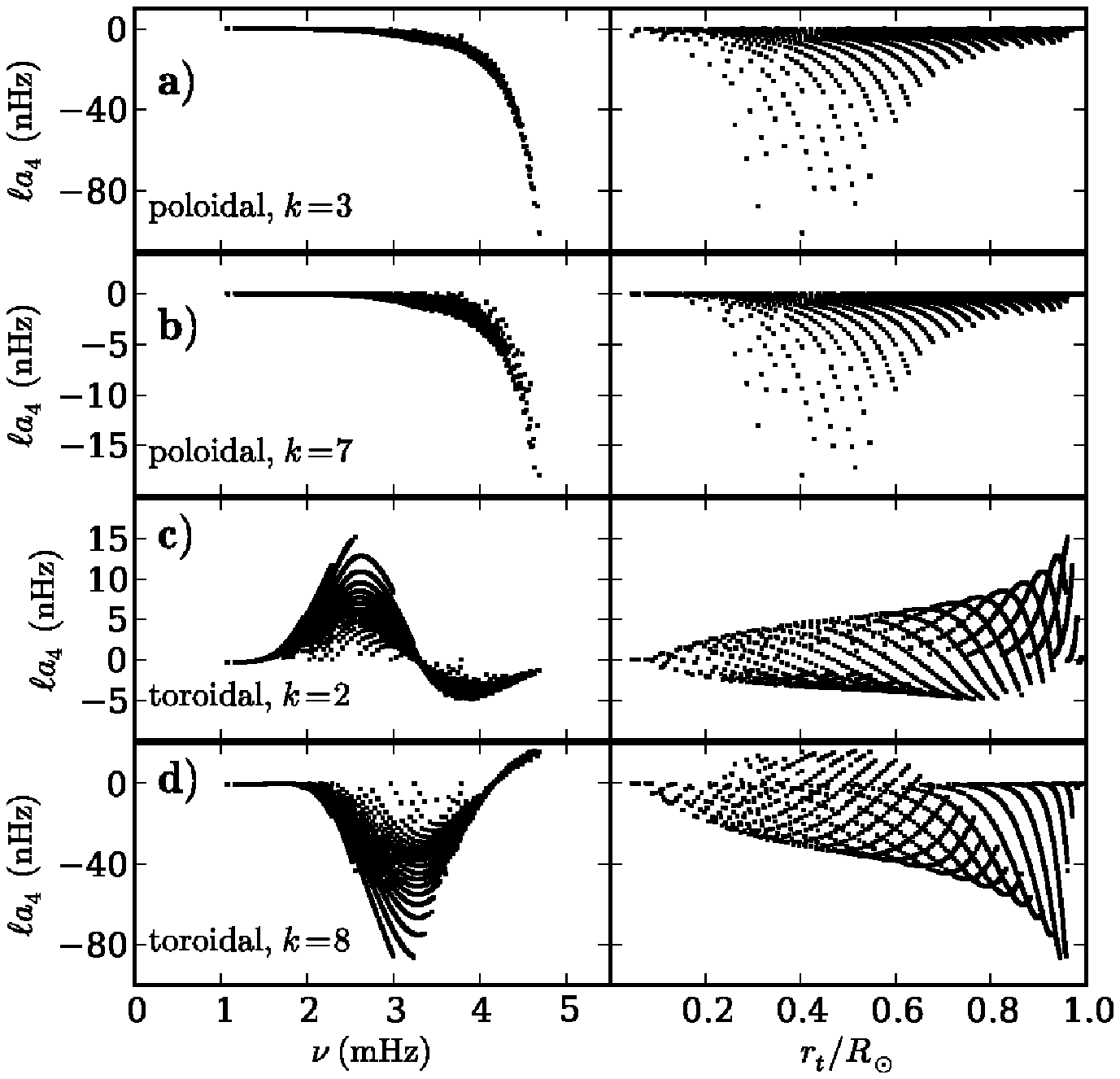}
\caption{Effects of various magnetic fields on the $\ell a_4$ splitting coefficients, 
as a function of both frequency $\nu$ (left hand panels) and lower turning 
radius $r_t$ (right hand panels).  The top two panels show the results for poloidal fields 
with $k=3$ (panel a) and $k=7$ (panel b).  The bottom two panels show the results for toroidal 
fields, both with $\beta_0=10^{-4}$, $r_0 = 0.999R_\odot$ and $d_0=0.001R_\odot$.  
Panel (c) is for a $k=2$ field and panel (d) for a $k=8$ field.  Only observed modes have been plotted.
\label{fig:model_a4}}
\end{figure}

The $a_4$ splitting coefficients due to various poloidal and toroidal 
fields are shown in Fig.~\ref{fig:model_a4}, which shows the results
for two poloidal fields, the $k=3$ and $k=7$ fields, as well as two toroidal fields 
with different values of $k$ ($k=2$ and $k=8$), each with $r_0=0.999R_\odot$ 
and $d_0=0.001R_\odot$.  The $k=1$ field has essentially no effect on the $a_4$ 
splitting coefficients.

\subsection{Fits to observed data}
In order to choose the fields which best match the actual data, we 
have computed the splitting coefficients for 
a large grid of field configurations, with fields throughout the 
convection zone.  For poloidal fields we varied $k$ --- the form 
of the radial profile was found not to matter very much for the 
splitting coefficients, so long as the field penetrated below the surface.  
For the toroidal fields, we varied the location $r_0$, the width of 
the field $d_0$, and the latitudinal distribution with $k$.  The 
range in $r_0$ was between $0.70R_\odot$ and $1.0R_\odot$.  The values 
for $d_0$ ranged from $10^{-4}R_\odot$ to $0.2R_\odot$.  In 
order to judge goodness-of-fit, we use the $\chi^2$ statistic.  For 
both the poloidal and the toroidal fields, the perturbations vary 
linearly with the square of the field strength, so to fit the field, 
we allowed the field strength to vary freely, and chose the strength 
that minimized the $\chi^2$.  We have computed the $\chi^2$ for all 
the field configurations in our grid, as well as for many combinations 
of two and three different fields.  The results we present below 
represent the best fits from the entire grid of computed models.

\begin{figure*}
\plotone{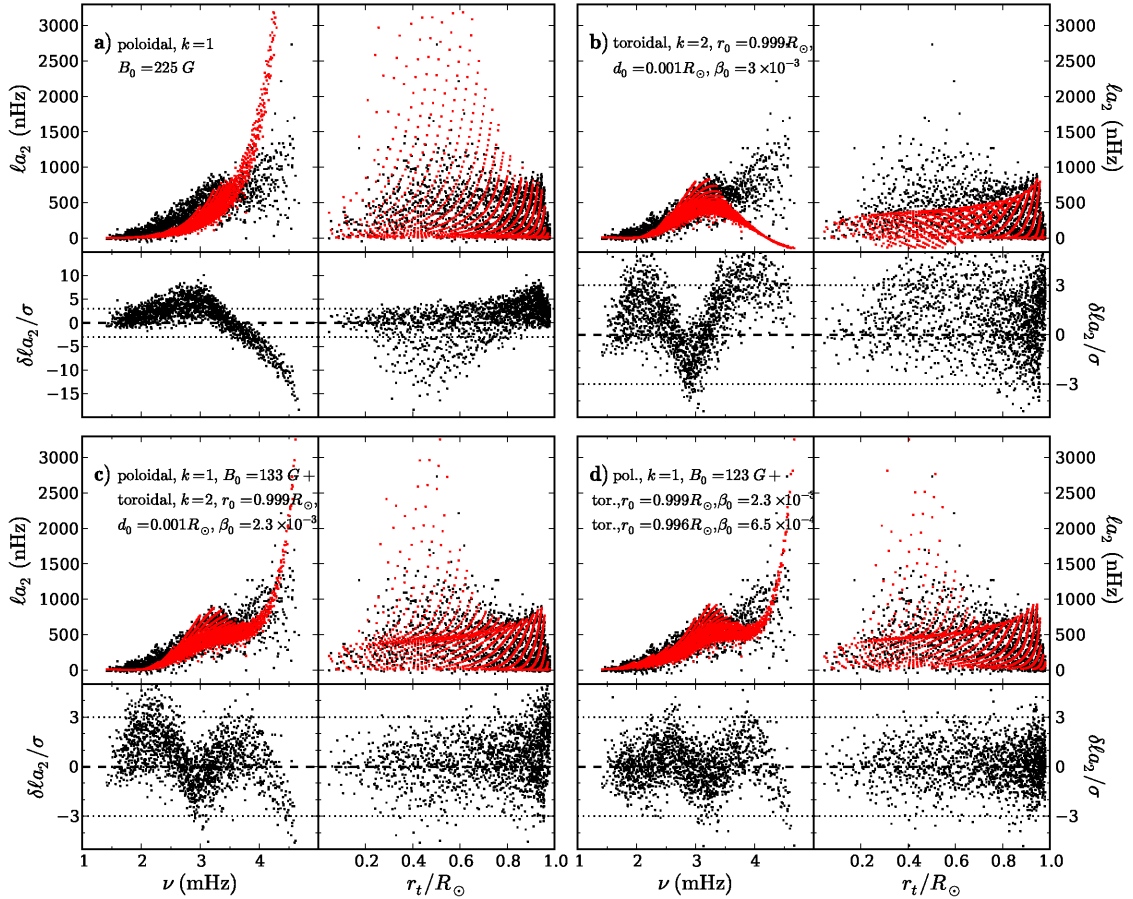}
\caption{Fits to observed splitting coefficients $\ell a_2$ for 
different magnetic field configurations.  Four different fits 
are shown, both as a function of frequency $\nu$ and as a function 
of lower turning radius $r_t$.  The data are shown in black, and 
the modeled points in red.  The residuals, scaled by the errors 
in the data, are also shown below the comparisons.  The data are the 
$\ell a_2$ splitting coefficients from an MDI 72 day mode parameter 
set, taken at the peak of solar cycle 23 (MDI set \#3160, start day 
2001 Aug 27).  Panel (a) shows a fit from a $k=1$ poloidal field.  
Panel (b) shows the fit from a near-surface toroidal field.  Panel (c) 
shows the best fit field combination with two fields to this data set --- 
a combination of a dipole poloidal field with a toroidal field 
located just below the surface ($r_0=0.999R_\odot$, $d_0=0.001R_\odot$).  
Panel (d) shows the best fit with three fields --- the same poloidal 
and toroidal field as in panel (c) (though with slightly different 
field strengths) and another toroidal field at $r_0=0.996R_\odot$ and a 
width of $d_0=0.001R_\odot$.
\label{fig:fit_high}}
\end{figure*}

The largest signal-to-noise ratio in $a_2$ is found at peak activity, and 
so the highest activity set ought to be the easiest to fit.  
Comparison of different field configurations with the splitting 
coefficients at high activity are shown in Fig.~\ref{fig:fit_high}, 
and the fits are shown both as a function of frequency and as a function 
of lower turning radius.  The residuals, normalized by the errors in
the data, are also shown.  A fit to a $k=1$ poloidal field is shown in 
panel (a).  The reduced $\chi^2$ for this fit is 16, and it is evident 
that the field does a poor job of reproducing the observed splitting 
coefficients.  Higher order poloidal fields are considerably worse, as 
an examination of Fig.~\ref{fig:model_pol} will show --- these fields 
perturb all the splitting coefficients negatively, whereas the observed 
splittings are all positive.  Panel (b) shows the effect of a toroidal field situated 
near the surface.  Although we attempted to fit toroidal fields throughout 
the convection zone, fields not located very near the surface were extremely 
poor fits to the data.  The field shown in panel (b) is the best fit for a 
single toroidal field, with $r_0=0.999R_\odot$ and $d_0=0.001R_\odot$.  The 
reduced $\chi^2$ is 5.  The residuals are mostly without structure in $r_t$, 
but are oscillatory in frequency, a  hint that there could be a second, somewhat
deeper field.  The splitting coefficients at peak solar activity cannot be well 
fit by either a toroidal field or a poloidal field of the form considered
by us.  

The third field configuration shown (panel c) is a combination of a $k=1$ 
poloidal field and a near-surface toroidal field ($r_0=0.999R_\odot$, 
$d_0 = 0.001R_\odot$ --- the same field from panel (b)).  This combination of 
fields yields a much better fit to this data set, with a reduced $\chi^2$ of 
2.8.  Using a surface toroidal field instead of a poloidal field does not 
fit the data as well, although it is an improvement over the fit in (b), with a 
$\chi^2$ value of 3.5.  
Like the toroidal-only fit, the residuals are more or less 
without structure in $r_t$, but show oscillatory behavior in frequency.  
The peak field strengths of the two fields are 133~G and 368~G for the 
poloidal and toroidal fields, respectively.  The residuals from this fit 
can be fit by a toroidal field centered at $r_0=0.996R_\odot$, so in 
panel (d) we show the best fit to the data:  a $k=1$ poloidal field with 
two toroidal fields, one centered at $r_0=0.999R_\odot$ and another centered 
at $r_0=0.996R_\odot$.  Both toroidal fields are $k=2$ fields and have 
widths $d_0=0.001R_\odot$.  The poloidal field has a peak field strength 
at the surface of $124\pm18$~G, while the toroidal fields have peak field 
strengths of $380\pm30$~G and $1.4\pm0.2$~kG, respectively.  The reduced 
$\chi^2$ of this fit is 1.7.  Attempts to fit the data with a single 
toroidal field which occupies the same region as the two field in this fit 
did not yield a good fit --- the data seem to require a double peaked 
field.

\begin{figure*}
\plotone{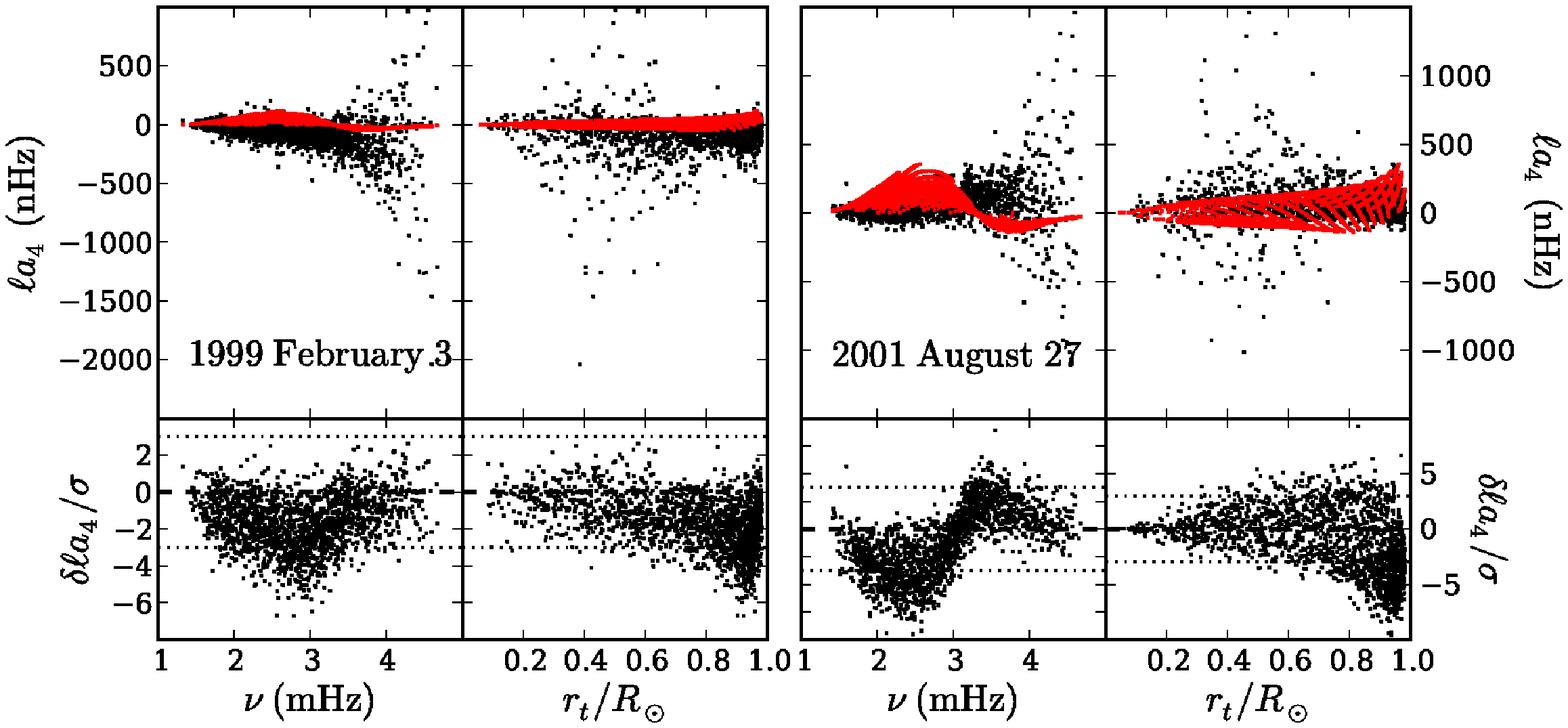}
\caption{Comparisons of data to models for $\ell a_4$ splitting 
coefficients from two different data sets.  The field configurations are 
the same as from Fig.~\ref{fig:fit_high} (a).  The left hand panels  
are from set \#2224, the right hand panels are from \#3160.  As 
in the previous figure, data and model are shown both as a 
function of frequency $\nu$ and as a function of lower turning radius 
$r_t$.  The residuals are shown below the data, and are normalized by 
the errors in the data. The model is obtained by fitting only the
$a_2$ coefficients.
\label{fig:fit_a2_a4}}
\end{figure*}

Figure \ref{fig:fit_a2_a4} shows the $a_4$ splitting coefficients
for two different data sets (\#2224 and \#3160), and the same models 
shown in Fig.~\ref{fig:fit_high} panel (d).  The 
errors on the data are large compared to the signal --- thus the 
normalized residuals are comparable to those for the $a_2$ 
coefficients, but some other field configurations also fit the $a_4$ 
coefficients equally well, so we do not use them to constrain the 
field configurations or determine the field strengths.

\begin{figure*}
\plotone{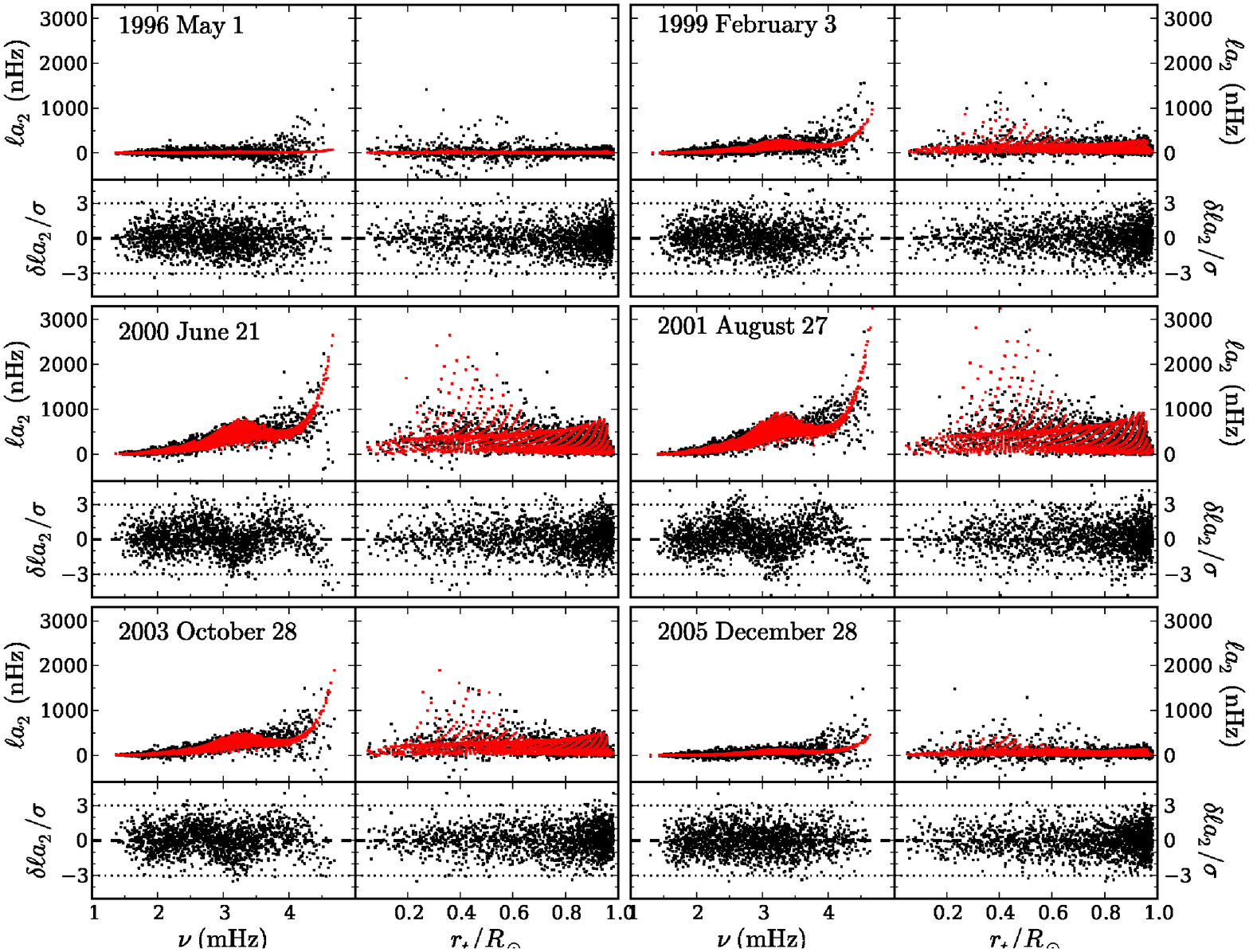}
\caption{Fits to measured splitting coefficients $\ell a_2$ for 
six different sets throughout solar cycle 23.  The data (shown in 
black) are from MDI 72 day mode parameter sets.  The magnetic 
field configuration is the same as panel {\bf d\/} in Figure 
\ref{fig:fit_high}:  a dipole poloidal field and two toroidal fields 
at $r_0=0.996R_\odot$ and $r_0=0.999R_\odot$, with $d_0=0.001R_\odot$ 
and $k=2$.  The fits are shown both as a function of frequency $\nu$ 
and as a function of lower turning radius $r_t$.  The residuals 
scaled by the errors in the data are also shown.  The toroidal field 
strengths at $r=0.999R_\odot$ correspond to $\beta_0$ = $10^{-4}$, $7.8\times10^{-4}$, 
$2\times10^{-3}$, $2.5\times10^{-3}$, $1.3\times10^{-3}$, and $5\times10^{-4}$ 
for the six sets, respectively.  The toroidal field strengths at $r=0.996R_\odot$ 
correspond to $\beta_0$ = $1.2\times 10^{-5}$, $2 \times 10^{-4}$, $5.6\times 10^{-4}$, 
$6.6\times 10^{-4}$, $2.9\times 10^{-4}$, and $1.4\times 10^{-4}$.  
The poloidal field strengths at the surface are $B=0$~G, 68~G, 115~G, 125~G, 
94~G, and 58~G.
\label{fig:fit_all}}
\end{figure*}

Having fit the high activity set, we repeat the fits for all 56 sets 
in our study.  Figure \ref{fig:fit_all} shows fits of the same $k=1$ poloidal plus 
toroidal field combination to six representative mode sets, covering 
the rise and fall of solar cycle 23.  The first set, \#1216, is the 
first 72-day mode set from the MDI program, and is a low activity set.  
Two rising phase sets are shown, \#2224 and \#2728, with 10.7~cm fluxes 
of 131.4~SFU and 187.3~SFU, respectively.  The high activity set from 
Fig.~\ref{fig:fit_high} is shown, and a declining phase set (\#3952, 
$F_\mathrm{10.7}=126.4$) and a set from the current minimum (\#4744).  
For all sets, the same combination of poloidal and toroidal fields  ---
but with differing field strengths --- was found to be the best fit.  
The residuals show the same structure with frequency.

No mode set in our study is well fit by any magnetic field at the 
base of the convection zone.  For low activity sets (10.7~cm flux of 
less than 100~SFU), we can find an upper bound on the field that 
could be present in the data.  For a field centered at the base of 
the convection zone with $k=2$ and $d_0=0.01R_\odot$, we find that a 
fields of up to 300~kG can be fit to the data, although this is an 
upper limit, not a detection, since the models of that field strength 
or lower give the same $\chi^2$ as a zero field strength model.  At 
high activity, the dominant signal is from the surface, which we have 
attempted to explain with magnetic fields located in those layers.

\begin{figure}
\plotone{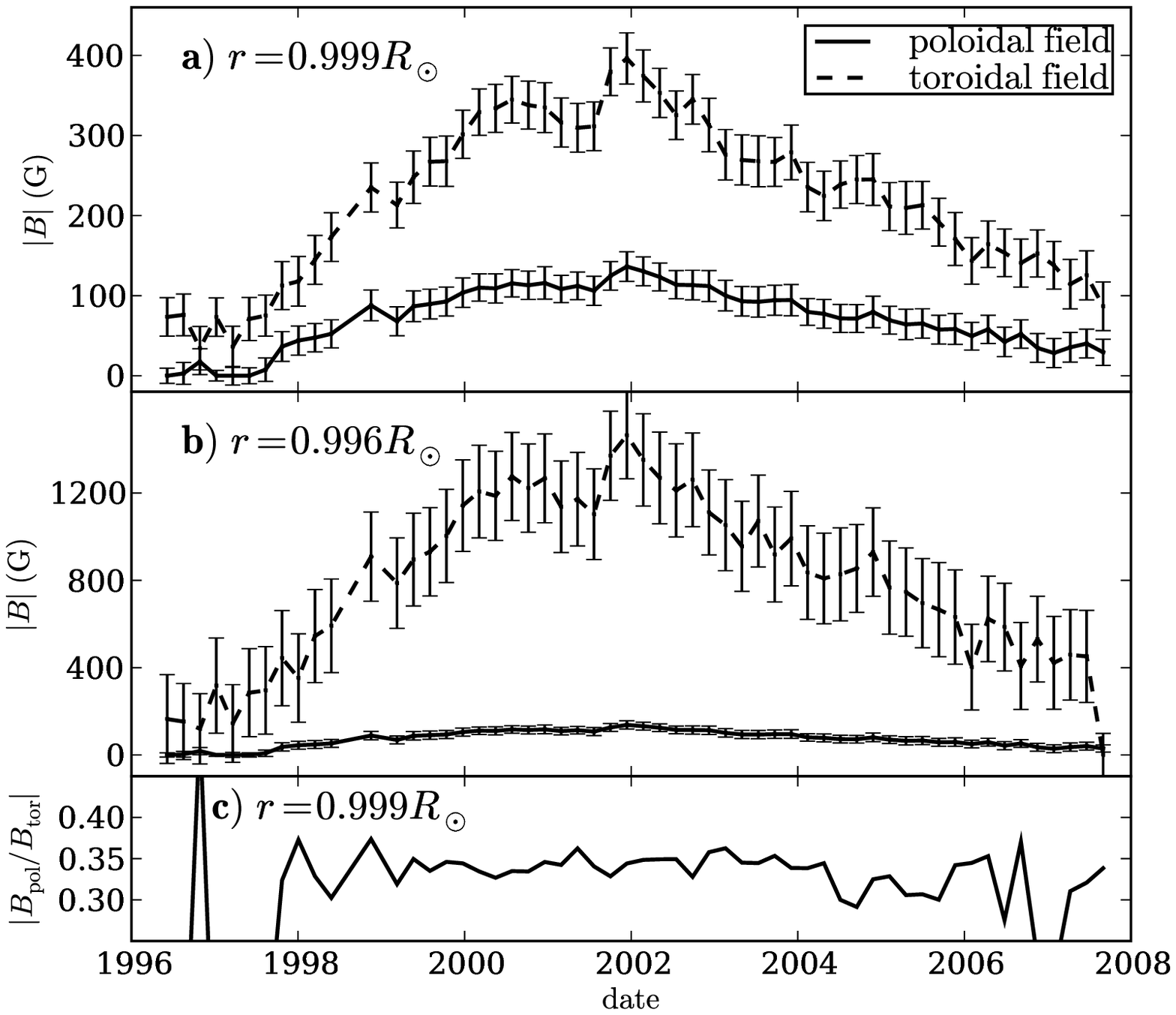}
\caption{The strength of the inferred magnetic fields as a function 
of time over solar cycle 23.  Each MDI 72 day mode parameter set is 
fitted by with the same magnetic field configuration as Fig.~\ref{fig:fit_all}.  
The strengths (in Gauss) of the poloidal field at $r=0.999R_\odot$ 
(solid black line) and the toroidal field at $r=0.999R_\odot$ 
(dashed line) are shown in the upper panel (a).  The middle panel (b) shows the same 
quantities as in the upper panel, but this time at a radius of $r=0.996R_\odot$.  
The lower panel (c) shows the ratio of the poloidal field strength to the toroidal 
field strength at $r=0.999R_\odot$.  The ratio of poloidal to toroidal at 
$r=0.996R_\odot$ looks very similar.
\label{fig:strength}}
\end{figure}

\begin{figure}
\plotone{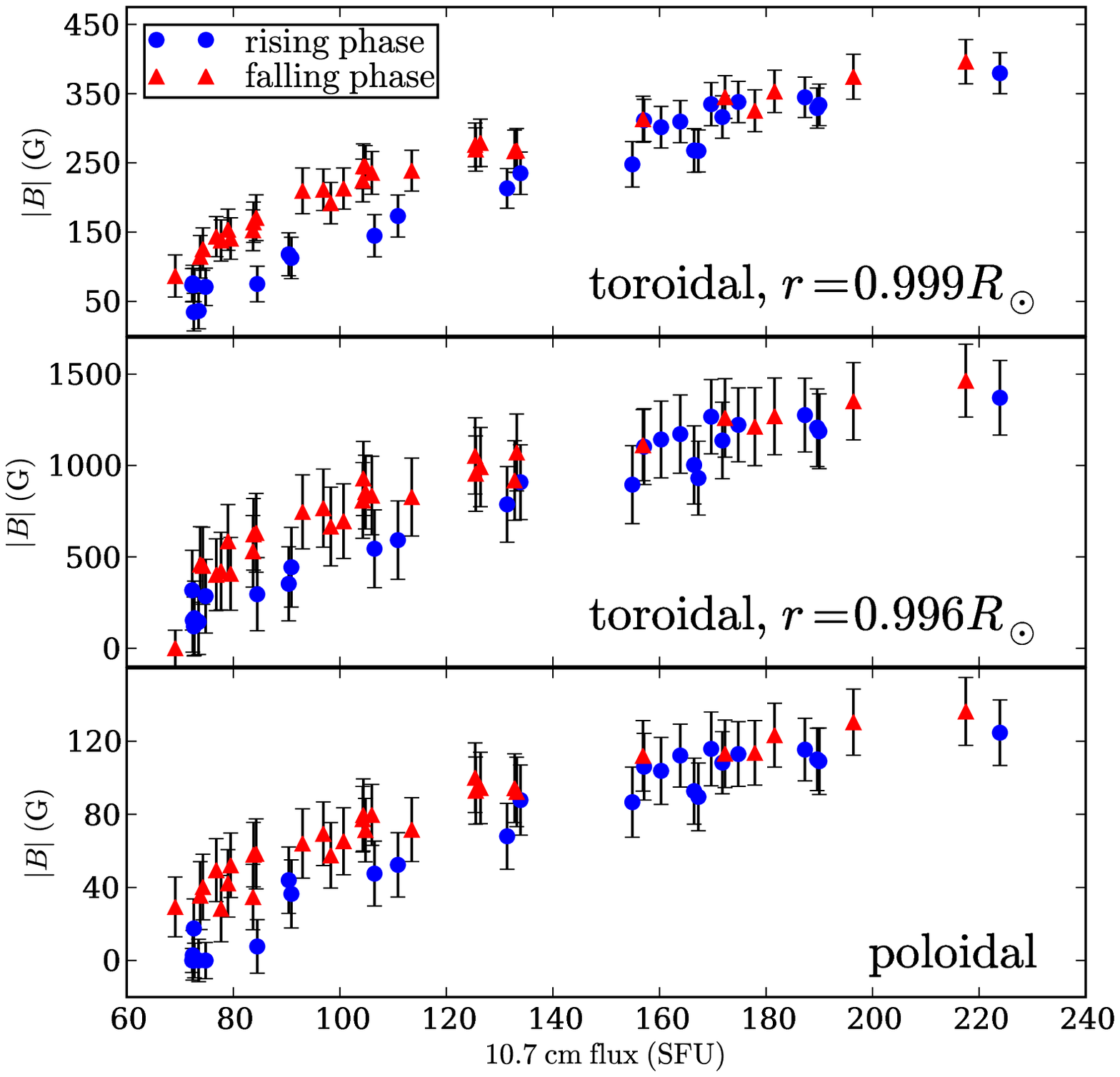}
\caption{The strength of the inferred magnetic fields as a function 
of 10.7 cm radio flux.  The top panel shows the $r_0=0.999R_\odot$ toroidal 
field strength, the middle panel shows the $r=0.996R_\odot$ toroidal 
field strength, and the bottom panel shows the poloidal field strength 
at $r=0.999R_\odot$.  Rising and declining phase are distinguished with 
blue circles for the rising phase and red triangles for the declining phase.  
The toroidal field shows a hysteresis effect.  The poloidal field shows 
some hysteresis at low activity, as well as a hint of saturation at high 
activity.
\label{fig:activity}}
\end{figure}

The field strengths of the poloidal and shallow toroidal field fits to 
all 56 sets used in this study are shown in Fig.~\ref{fig:strength}.  
Also shown in this figure is the ratio of the poloidal field strength 
to the $r_0=0.999R_\odot$ toroidal field strength.  With the exception 
of the low activity sets at the beginning and end of the solar cycle, 
where the uncertainty in the fits is relatively large, the ratio between 
the poloidal and toroidal field strengths is roughly constant.  The 
field strengths from Fig.~\ref{fig:strength} are correlated with global 
activity indices from solar cycle 23.  The correlation coefficients are 
0.90, 0.93, and 0.92 for the poloidal and two toroidal field components, 
respectively.  In Fig.~\ref{fig:activity}, we plot the toroidal and 
poloidal field strengths as a function of one such global index, the 
10.7~cm radio flux.  The field strengths prove to be highly correlated 
with activity, although there is a hysteresis-like effect evident in 
the toroidal field strengths --- the rising phase (shown in blue) is 
weaker than the declining phase fields.  The same effect may also be 
present at low activity in the poloidal field strengths.  The poloidal 
field strengths do seem to saturate at high activity. The strengths of 
the two toroidal fields are extremely well correlated.

\section{Discussions and conclusions}
We have attempted to use the first even order splitting coefficient ($a_2$) 
to infer the configuration and strength of the Sun's internal magnetic 
fields over the course of solar cycle 23, assuming that the entire 
signature in $a_2$ after correction for rotation effects is magnetic 
and that the fields are axisymmetric.  The field that we have found is 
a combination of poloidal field and a double-peaked near-surface toroidal 
field. The strengths of the poloidal and toroidal components, at least for 
high activity period, are well correlated.  The relative strengths of the 
two toroidal fields are also extremely well correlated.  

Although the fits we have shown are the best fit to the data from the 
grid of models that we have computed, we can say nothing about the uniqueness 
of these fits over the set of all possible magnetic field configurations 
in the solar interior.  In particular, the choice of radial profile of the 
toroidal fields is virtually limitless, and by restricting our work to 
profiles of the form (\ref{tor}), we have limited our search to a restricted
class of fields. It is possible that there are fields we did not 
consider with quite different radial and latitudinal distributions which 
fit the data as well as the fields we have presented as best fits.  In 
addition, as noted above, it is not strictly correct to add the splitting 
coefficient perturbations together as we have done without explicitly 
accounting for the perturbations arising from the cross terms.  We do 
not expect, however, these corrections to be significant, and a full 
treatment of these corrections would be considered in a future work.

Our inferred magnetic field does not change its latitudinal distribution 
over the course of the solar cycle.  This is in part due to the fact that 
we are only fitting the $a_2$ coefficient (as noted above, the higher order 
splittings had large errors), so our sampling of the interior is not really 
latitudinally sensitive.  Thus, we do not see a butterfly diagram in our 
magnetic fields.  \citet{UB05} measured the surface toroidal component of 
the solar magnetic field over almost an entire 22-year cycle.  The field 
they measure is roughly a tenth of the peak strength of our toroidal field.  Strictly 
speaking, however, we see no toroidal field at all at the surface, since in 
our inferred field, the field strength becomes zero precisely at $r=1R_\odot$.  
The peak strength that we measure, however, is only 700~km below the surface, 
and the field could penetrate the surface somewhat.  \citet{UB05} find a field 
which gives a $\beta \sim 6 \times 10^{-5}$ at the surface at high activity, 
and drops to nothing at low activity, while we find a field that changes from 
$\beta_0 \sim 10^{-4}$ at low activity to $\beta_0=2\times10^{-3}$ at high
activity at a radius of $r=0.999R_\odot$ (a depth of approximately 700~km).

Recently, attention has been focused on the strength and configuration of
the quiet Sun surface magnetic field.  \citet{Harveyetal07} reported the
presence of a `seething' horizontal magnetic field with an rms field
strength of 1.7~G.  With the launch of Hinode (Solar B), the high spatial
resolution of the onboard spectropolarimeter has been used to study the
horizontal fields of the solar photosphere.  \citet{Lites07,Lites08}
have measured the horizontal flux, which they find to be 55~G, compared
to the average vertical flux of 11~G.  \citet{PP09} found that the zonal
component (component in the East-West direction) was much smaller than
the radial component, reporting an inclination angle of less than $12^\circ$
from vertical in the East-West direction.

The fields being studied in the aforementioned works are generally very 
tangled fields which thread through the intergranular lanes and so 
they are not axisymmetric fields.  It is worthwhile to compare 
our results with theirs, since tangled fields on local scales can organize 
into roughly axisymmetric fields on global scales.
However, the contribution to splittings are more sensitive to $\langle B^2\rangle$ rather than
$\langle B\rangle^2$ and hence tangled field may also contribute to it, even
when the average $\langle B\rangle$ is very small.
Further, considering the general behavior of the perturbation to the mode frequencies,
our inference about the location of required magnetic field is more robust
as a different location will yield a very different behavior of splitting
coefficients. The exact magnitude of the field may depend on the assumption
of geometry and on it being tangled or large scale. Nevertheless, we
believe that our estimate is of the right order, though the statistical errorbars
obtained by us may not be realistic. The systematic errors in these
estimates would be certainly larger.  The  
dominance of poloidal field orientation at the surface found by \citet{PP09} 
is found in our own results --- at the surface, the toroidal field is 
weak or vanishing, but the poloidal field remains.  In the period analyzed 
by \citet{Lites07,Lites08}, we find a poloidal field strength of 40~G, and 
a toroidal field of 90~G at a depth of 700~km.  The vertical flux they 
find (11~G) is weaker than what we detect, but their 55~G horizontal 
flux may be roughly consistent with our toroidal field.

\citet{SL08} found that the dipole moment of the surface magnetic field, 
measured from MDI magnetograms, was half the strength in 2008 that it was 
in 1997, during the last solar minimum.  We do not see such a difference 
from the beginning of our period to the end --- in fact, we find the 
poloidal field strength is slightly higher during the current minimum, 
although the level of the difference is within the errors, and our data 
sets end in 2007, so the comparison is not contemporaneous.

Hysteresis in the relations between activity indices has been observed 
before, for example in the relation between low degree \citep{AGetal92,J-Retal98} 
and intermediate degree \citep{Tripathyetal00,Tripathyetal01} acoustic modes 
and global magnetic indices.  It should be noted that an analysis of a 
full solar cycle's worth of intermediate degree $p$-modes data does not show
any hysteresis in mean frequencies as a function of 10.7~cm flux \citep{BB08}.  
\citet{Tripathyetal01} noted that, among the global mode indices, the 
relation between global line-of-sight magnetic flux and 10.7~cm radio flux 
showed a hysteresis effect, but the relation between the radiative indices 
and 10.7~cm flux did not.  \citet{M-IS00} argued that the observed hysteresis 
could be almost entirely due to the latitudinal distribution of magnetic flux 
on the surface of the Sun.  We believe that this is a compelling explanation 
for the hysteresis that we find.  The 10.7~cm flux is the integrated flux 
received at the Earth and does not contain any information about the 
latitudinal variation, while the $a_2$ splitting coefficient is associated 
with definite latitudinal variation, given by $P_2(\cos\theta)$ \citep{ABH01},
and hence the two would not be the same.  More importantly, we expect the 
actual magnetic fields in the near surface layers to drift equatorward --- as 
the surface fields do.  

Few conclusions can really be drawn from this work with respect to 
dynamo theory since the fields we have inferred are predominantly shallow fields, 
whereas most dynamo models operate much deeper down, in the shear layer 
at and below the base of the convection zone.  \citep[some useful 
recent reviews include][]{Oss2003,Char2005,MT2009}.  The upper limits that 
we place on fields at that depth are consistent with earlier helioseismic 
results \citep[e.g.,][]{Basu97, ACT00, CS02,CS05,BB08}.  Many deep-seated dynamo 
mechanisms predict an anticorrelation between the poloidal and toroidal 
field components, as the dynamo converts poloidal to toroidal field and 
toroidal field back to poloidal.  We do not see any evidence of such 
conversion.  Some dynamo mechanisms, however, operate in the near-surface 
shear layer \citep[e.g.][]{Brand05}.  Although the fields generated in 
these models are generally extremely tangled, on global scales these 
fields can have toroidal and poloidal components \citep[e.g.][]{BBBMNT07,BBMBT09}.  
In particular, although they were considering a more rapidly rotating 
star than the Sun, \citet{BBBMNT07} noted that their field contained 
both a poloidal and a toroidal component, and that the toroidal component 
was much the stronger of the two.

\acknowledgements
This work utilizes data from the Solar Oscillations Investigation/ 
Michelson Doppler Imager (SOI/MDI) on the Solar and Heliospheric 
Observatory (SOHO).  SOHO is a project of international cooperation 
between ESA and NASA.  MDI is supported by NASA grants NAG5-8878 
and NAG5-10483 to Stanford University.  This work was partially supported by NSF 
grants ATM 0348837 and ATM 0737770 to SB.  CB is supported by a NASA 
Earth and Space Sciences Fellowship NNX08AY41H.


\begin{thebibliography}{}
\bibitem[Antia \& Basu(1994)]{AB94} Antia, H.~M., \& Basu, S.\ 1994, \aaps, 107, 421

\bibitem[Antia et al.(2001)]{ABH01} Antia, H.~M., Basu, S., Hill, F.,
Howe, R., Komm, R. W., \& Schou, J. 2001, MNRAS, 327, 1029

\bibitem[Antia et al.(2008)]{ABC08} Antia, H.~M., Basu, S., 
\& Chitre, S.~M.\ 2008, \apj, 681, 680

\bibitem[Antia et al.(2000)]{ACT00} Antia, H.~M., Chitre, S.~M., \& 
Thompson, M.~J.\ 2000, \aap, 360, 335 

\bibitem[Baldner \& Basu(2008)]{BB08} Baldner, C.~S., \& Basu, S.\ 2008, \apj, 686, 1349 

\bibitem[Baldner et al.(2008)]{BBL09} Baldner, C.~S., Basu, S., \& Larson, T.~P.\ 2008, 
in ASP Conf.~Ser.~GONG 2008/SOHO XXI Meeting on Solar-Stellar Dynamos as Revealed by 
Helio-and Asteroseismology, ed. Dikpati, M., Gonzalez-Hernandez, I., Arentoft, T., \& Hill, F., 
in press

\bibitem[Basu et al.(1994)]{BAN94} Basu, S., Antia, H.~M., 
\& Narasimha, D.\ 1994, \mnras, 267, 209

\bibitem[Basu(1997)]{Basu97} Basu, S.\ 1997, \mnras, 288, 572

\bibitem[Brandenburg(2005)]{Brand05} Brandenburg, A.\ 2005, \apj, 625, 539

\bibitem[Brown et al.(2007)]{BBBMNT07} Brown, B.~P., Browning, M.~K., Brun, A.~S., 
Miesch, M.~S., Nelson, N.~J., \& Toomre, J.\ 2007, in AIP Conf.~Ser.~948, Unsolved 
Problems in Stellar Physics: A Conference in Honor of Douglas Gough, ed. Stancliffe, R.~J.,
 Houdek, G., Martin, R.~G., \& Tout, C.~A., 271

\bibitem[Brown et al.(2009)]{BBMBT09} Brown, B.~P., Browning, 
M.~K., Miesch, M.~S., Brun, A.~S., \& Toomre, J.\ 2009, arXiv:0906.2407

\bibitem[Charbonneau(2005)]{Char2005} Charbonneau, P.\ 2005, Living Reviews in Solar 
Physics, 2, 2

\bibitem[Chou \& Serebryanskiy(2002)]{CS02} Chou, D.-Y., \& Serebryanskiy, A.\ 2002, 
\apjl, 578, L157 

\bibitem[Chou \& Serebryanskiy(2005)]{CS05} Chou, D.-Y., \& Serebryanskiy, A.\ 2005, 
\apj, 624, 420

\bibitem[D'Silva \& Choudhuri(1993)]{DC93}
D'Silva, S. \& Choudhuri, A. R. 1993, A\&A, 272, 621

\bibitem[Dziembowski \& Goode(1984)]{DG84} Dziembowski, W., \& Goode, P.~R.\ 1984, 
Memorie della Societa Astronomica Italiana, 55, 185 

\bibitem[Dziembowski \& Goode(1988)]{DG88} Dziembowski, W., \& Goode, P.~R.\ 1988, 
in IAU Symp. 123, Advances in Helio- and Asteroseismology, ed. Christensen-Dalsgaard, 
J.~C.~D. \& Frandsen, S., 171 

\bibitem[Dziembowski et al.(2000)]{DGKS00} Dziembowski, W.~A., 
Goode, P.~R., Kosovichev, A.~G., \& Schou, J.\ 2000, \apj, 537, 1026 

\bibitem[Dziembowski et al.(2001)]{DGS01} Dziembowski, W.~A., 
Goode, P.~R., \& Schou, J.\ 2001, \apj, 553, 897

\bibitem[Dziembowski et al.(1990)]{Dzetal90} Dziembowski, W.~A., 
Pamyatnykh, A.~A., \& Sienkiewicz, R.\ 1990, \mnras, 244, 542

\bibitem[Gough(1990)]{G1990} Gough, D.~O.\ 1990, Lecture Notes in Physics, 367, 
Progress of Seismology of the Sun and Stars, ed. Osaki, Y.~ \& Shibahashi, H., 283

\bibitem[Gough \& Thompson(1990)]{GT90}Gough, D.~O., \& 
Thompson, M.~J.\ 1990, \mnras, 242, 25 

\bibitem[Anguera Gubau et al.(1992)]{AGetal92} Anguera Gubau, M., Palle, 
P.~L., Perez Hernandez, F., Regulo, C., \& Roca Cortes, T.\ 1992, \aap, 255, 363

\bibitem[Harvey et al.(2007)]{Harveyetal07} Harvey, J.~W., Branston, 
D., Henney, C.~J., \& Keller, C.~U.\ 2007, \apjl, 659, L177

\bibitem[Isaak(1982)]{Isaak82} Isaak, G.~R.\ 1982, \nat, 296, 130 

\bibitem[Jimenez-Reyes et al.(1998)]{J-Retal98} Jimenez-Reyes, S.~J., Regulo, C., 
Palle, P.~L., \& Roca Cortes, T.\ 1998, \aap, 329, 1119

\bibitem[Larson \& Schou(2008)]{LarsonSchou08} Larson, T.~P., \& Schou, J.\ 2008, 
Journal of Physics Conference Series, 118, 012083

\bibitem[Lites et al.(2008)]{Lites08} Lites, B.~W., et al.\ 
2008, \apj, 672, 1237

\bibitem[Lites et al.(2007)]{Lites07} Lites, B., et al.\ 2007, 
\pasj, 59, 571

\bibitem[Miesch \& Toomre(2009)]{MT2009} Miesch, M.~S., \& Toomre, J.\ 2009, 
Annual Review of Fluid Mechanics, 41, 317

\bibitem[Moreno-Insertis \& Solanki(2000)]{M-IS00} Moreno-Insertis, F., \& 
Solanki, S.~K.\ 2000, \mnras, 313, 411

\bibitem[Ossendrijver(2003)]{Oss2003} Ossendrijver, M.\ 2003, in ASP Conf.~Ser.~ 286, 
Current Theoretical Models and Future High Resolution Solar Observations: 
Preparing for ATST, ed. Pevtsov, A.~A. \& Uitenbroek, H., 97

\bibitem[Petrie \& Patrikeeva(2009)]{PP09} Petrie, G.~J.~D., 
\& Patrikeeva, I.\ 2009, \apj, 699, 871

\bibitem[Ritzwoller \& Lavely(1991)]{RL91} Ritzwoller,
M.~H., \& Lavely, E.~M.\ 1991, \apj, 369, 557

\bibitem[Roxburgh \& Vorontsov(1994)]{RV94} Roxburgh, I.~W., 
\& Vorontsov, S.~V.\ 1994, \mnras, 268, 880

\bibitem[Schou et al.(1998)]{Schouetal98} Schou, J., et al.\ 1998, 
\apj, 505, 390 

\bibitem[Schou(1999)]{Schou99} Schou, J.\ 1999, \apjl, 523,
L181

\bibitem[Schrijver \& Liu(2008)]{SL08} Schrijver, C.~J., \& Liu, Y.\ 2008, 
\solphys, 252, 19

\bibitem[Thompson et al.(1996)]{Thompsonetal96} Thompson, M.~J., et 
al.\ 1996, Science, 272, 1300

\bibitem[Tripathy et al.(2000)]{Tripathyetal00} Tripathy, S.~C., 
Kumar, B., Jain, K., \& Bhatnagar, A.\ 2000, Journal of Astrophysics 
and Astronomy, 21, 357

\bibitem[Tripathy et al.(2001)]{Tripathyetal01} Tripathy, S.~C., 
Kumar, B., Jain, K., \& Bhatnagar, A.\ 2001, \solphys, 200, 3

\bibitem[Ulrich \& Boyden(2005)]{UB05} Ulrich, R.~K., \& Boyden, J.~E.\ 2005, 
\apjl, 620, L123

\bibitem[Zweibel \& Gough(1995)]{ZG95} Zweibel, E.~G., \& Gough, D.\ 1995, 
in ESA SP 376, Helioseismology, ed. Hoeksema, J.~T., Domingo, V., Fleck, B., 
\& Battrick, B., 73 


\end{thebibliography}
\end{document}